\documentclass[onecolumn,showpacs,preprintnumbers,amsmath,amssymb]{revtex4}
\usepackage{graphicx}
\usepackage{dcolumn}
\usepackage{bm}

\newcommand{\bef}{\begin{figure*}}
\newcommand{\eef}{\end{figure*}}

\newcommand{\be}{\begin{equation}}
\newcommand{\ee}{\end{equation}}
\newcommand{\bea}{\begin{eqnarray}}
\newcommand{\eea}{\end{eqnarray}}
\begin{document}

\title{Insight from the elliptic flow of identified hadrons measured in relativistic heavy-ion collisions}

\author{Praphupada Dixit and Md. Nasim}
\address{Department of Physics, Indian Institute of Science Education and Research, Berhampur 760010, India}

\begin{abstract}

In this paper, we have discussed the transverse momentum dependence of elliptic flow measured in relativistic heavy-ion collisions. Under the assumption of quark recombination model, $v_{2}$ of light and strange quarks are obtained from the measured $v_{2}$ of identified hadrons. The $v_{2}$ of strange quarks obtained from $\phi$-meson $v_{2}$ is found to be similar to that of from $\Omega$ $v_{2}$ in Au+Au collisions at $\sqrt{s_{NN}}$ = 200 GeV. This may indicates that both $\phi$ and $\Omega$ are produced through quark recombination at top RHIC energy. However, we find the $v_{2}$ of light quarks is consistently higher than that of strange quarks at the low transverse momentum. We have shown, $v_{2}$ of light and strange quarks  are consistent with each other when plotted as a function of its transverse kinetic energy. The observed scaling of $v_{2}$, works at both RHIC and LHC energies. It is also shown that the observed scaling holds for other flow harmonics, e.g. $v_{3}$. The new observed scaling is purely empirical and needs more theoretical input to understand the physics behind it.

\end{abstract}
\pacs{25.75.Ld}
\maketitle

\section{Introduction}
Quarks are the fundamental particles or building blocks of our universe. The bound states of quarks are known as hadrons. Due to strong coupling through gluon exchange, quarks are always found to be bound within hadrons. No free quarks exist in nature. The theory of quantum chromodynamics (QCD) predicts that at very high temperature and/or high density, hadrons might get melted into a deconfined matter of quarks and gluons. The deconfined medium of quarks and gluons is known as quark-gluon plasma (QGP). The relativistic heavy-ion collisions provide a unique opportunity to study the QGP in laboratory experiments~\cite{white}.
The medium created in the heavy-ion collision is very hot ($\sim$4 trillion degrees Celsius) and dense and also extremely short-lived ($\sim$10$^{-23}$ sec). In experiments, we are only able to detect the freely streaming final state particles emerging from the collisions. Using the information carried by these particles as probes, we can study the properties of the medium created in the collision.
The properties of the hot and dense medium formed in the heavy-ion collisions can be studied by using azimuthal  anisotropies of produced hadrons in momentum space. These momentum anisotropies  arise due to initial spatial anisotropies followed by multiple partonic interactions.  The azimuthal distribution of the produced particles can be described by using a Fourier series. Then, the Fourier coefficients of different harmonics quantify the degree of final state anisotropies. Within a hydrodynamical framework, the $v_{2}$ has been
shown to be sensitive to the equation of the state of the system formed in the 
collisions~\cite{flow1,flow2,hydro_flow}.\par
The elliptic flow parameter is defined as the $2^{\mathrm {nd}}$ Fourier coefficient, $v_{2}$, of  the particle distributions in emission azimuthal angle ($\phi$) with respect to
the reaction plane angle ($\Psi$)~\cite{art}, and can be written as
\begin{equation} \frac{dN}{d(\phi-\Psi)} \propto
1+2 v_2\cos(2(\phi - \Psi)). \end{equation}
For a given rapidity window, the second coefficient is
\begin{equation}
v_{2}=\langle\cos(2(\phi-\Psi))\rangle.
\end{equation}
The angular bracket represents the average over all the produced particles. Experimentally $v_{2}$ of different identified hadrons are measured as a function of transverse momentum $p_{T}$, centrality, and rapidity~\cite{pidv2_bes1,pidprl_bes1,pidprl_200,rhicflow,phobosflow,starflow,phenixflow}.\par
In this paper, we have discussed $v_{2}(p_{T})$ of identified hadrons measured in heavy-ion collisions and its physics implications.

\section{Results}
Experimentally, $v_{2}$ of different identified hadrons are measured as a function of transverse momentum at Relativistic Heavy-Ion Collider (RHIC) and Large Hadron Collider (LHC).
The measurements show a mass ordering in $v_{2}$ for $p_{T}$ $<$ 2 GeV/c and particle type dependence (i.e baryon vs. meson) at the intermediate $p_{T}$ (2.0 $<$ $p_{T}$ $<$ 4.0 GeV/c). The observed mass ordering is the effect of radial flow of the medium created in the collision~\cite{flow2,hydro_flow} and particle-type dependence can be explained by the quarks coalescence/recombination models~\cite{coal_fries,coal_greco,coal_voloshin,coal_all_1,coal_all_2,coal_all_3,coal_all_4}.\\

Within the framework of the coalescence mechanism, the $v_{2}$ of hadrons and quarks can be related by the following equation:
\begin{equation}
 v_{2}(p_{T})=n_{q}v_{2}^{q}(p_{T}/n_{q}), 
\label{eq1}
\end{equation}
where $p_{T}$ is the transverse momentum of hadrons, $n_{q}$ is the number of constituent quarks~\cite{coal_fries}. The $v_{2}$ and $v_{2}^{q}$ are the elliptic flow of hadrons and quarks, respectively.

                                                          
\bef
\begin{center}
\includegraphics[scale=0.8]{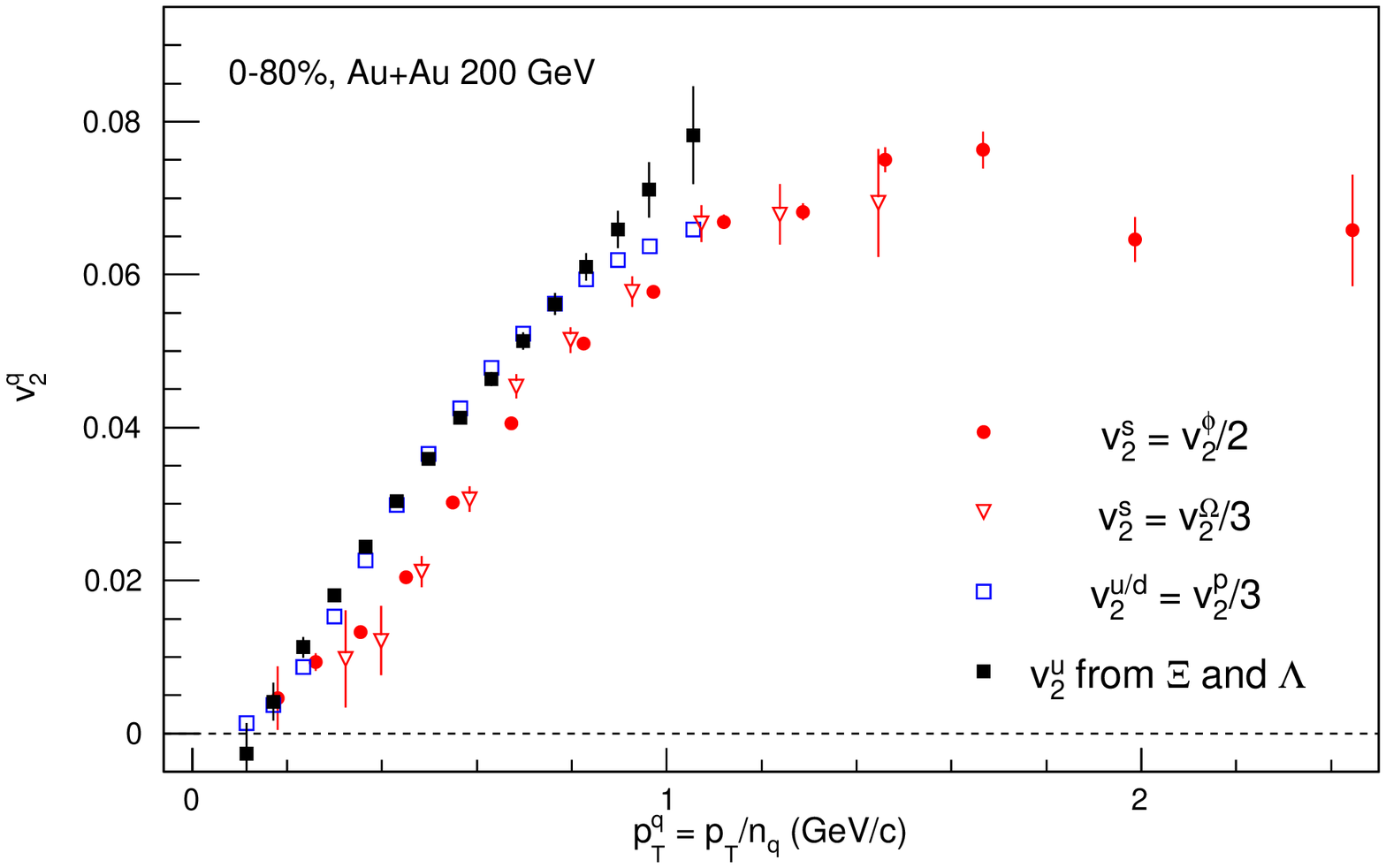}
\caption{(Color online) The  $v_{2}$ of light and strange quarks obtained from proton, $\phi$, $\Omega$, $\Lambda$, and $\Xi$ as a function of $p_{T}/n_{q}$ in Au+Au collisions at $\sqrt{s_{NN}}$ = 200 GeV~\cite{pidprl_200}.   }
\label{fig1}
\end{center}
\eef

Figure~\ref{fig1} shows the $n_{q}$ scaled $v_{2}$ as a function of $p_{T}/n_{q}$ in Au+Au collisions at $\sqrt{s_{NN}}$ = 200 GeV~\cite{coal_review}.  We observe that $n_{q}$ scaled $\phi$ $v_{2}$ is similar to $n_{q}$ scaled $\Omega$ $v_{2}$ within the measurement uncertainty. According to Ref.~\cite{phi_omega_star} the production of $\phi$ and $\Omega$ is mainly through the coalescence of strange quarks at these energies, therefore  both $n_{q}$ scaled $\phi$  and $\Omega$ $v_{2}$ should give $v_{2}$ of strange quarks and they should match. Our observation is consistent with the picture of $\phi$ and $\Omega$ production through quark coalescence.

Now,  $n_{q}$ scaled proton $v_{2}$ should be equivalent to $v_{2}$ of light quarks (considering up and down quarks have the same mass). We have also calculated $v_{2}$ of light quarks from the measured $v_{2}$  of $\Xi$ and $\Lambda$ using the following equation~\cite{coal_all_4,coal_all_5},
\begin{equation}
 v_{2}^{u}(p_{T})=\frac{1}{3}[2v_{2}^{\Lambda}((2+r)p_{T}) -v_{2}^{\Xi} ((1+2r)p_{T})] ;
\label{eq2}
\end{equation} 
where the factor $r = m_{s}/m_{u} $. Using constituent quark masses, the factor $r$ becomes 1.667.  We can see from Fig.~\ref{fig1}, the light quarks $v_{2}$ at 200 GeV obtained using Eq.~\ref{eq2} is equivalent to that obtained from the proton $v_{2}$.
From Fig.~\ref{fig1}, we see $n_{q}$ scaled proton $v_{2}$ is higher than $n_{q}$ scaled $\phi$  and $\Omega$ $v_{2}$ at low $p_{T}$. This could mean $v_{2}$ of strange quarks is smaller than $v_{2}$ of light quarks at low $p_{T}$ or this could be due to small hadronic interaction cross-sections of $\phi$  and $\Omega$ compared to non-strange hadrons~\cite{pidprl_200,refree_1,refree_3}. Alternatively, the difference between $n_{q}$ scaled proton $v_{2}$ and $n_{q}$ scaled $\phi$ $v_{2}$ at low $p_{T}$ could be due to the effect of radial flow on quarks $v_{2}$ at the partonic stage. 

\bef
\begin{center}
\includegraphics[scale=0.8]{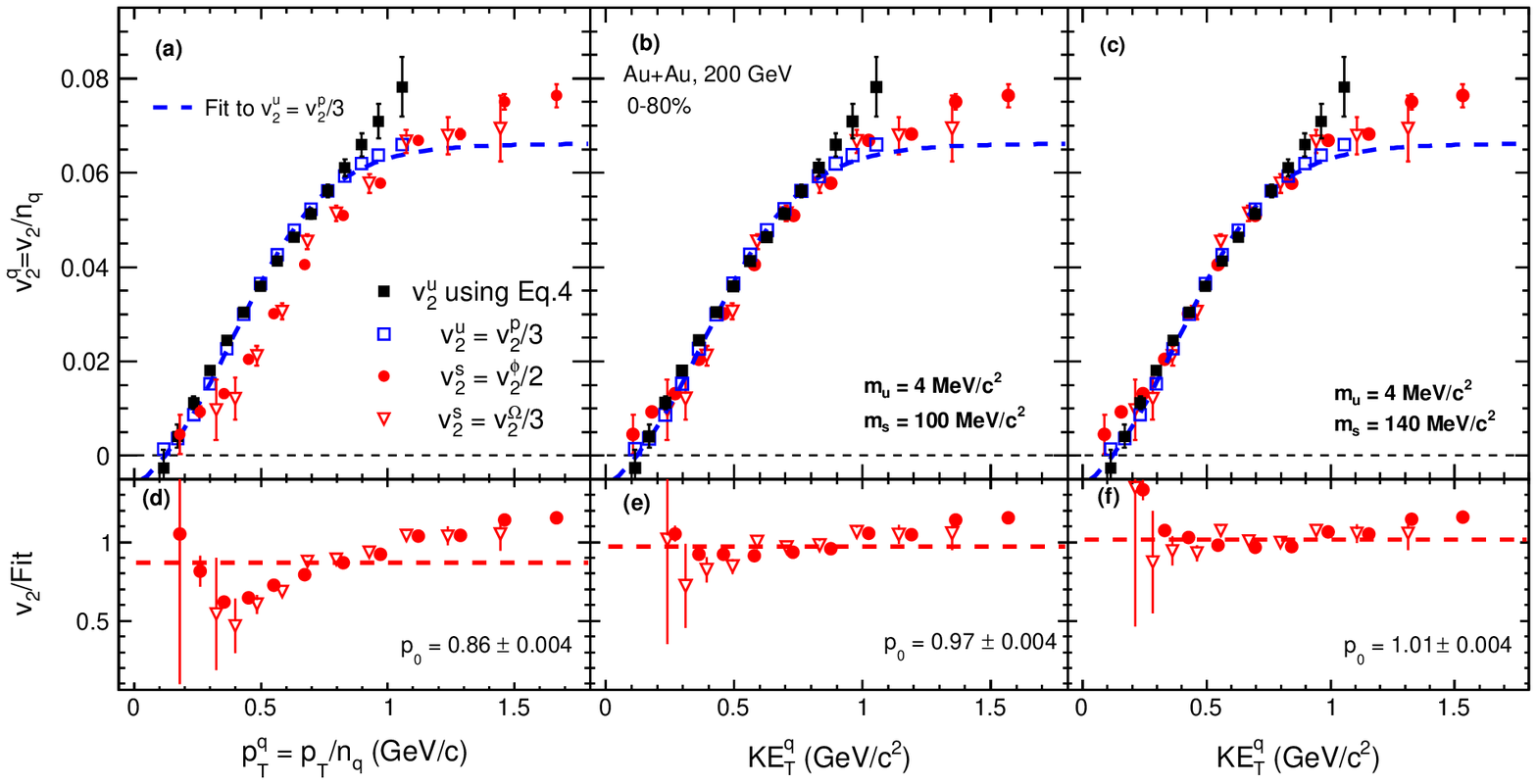}
\caption{(Color online) Panel (a) : The  $v_{2}$ of light and strange quarks obtained from proton, $\phi$, $\Omega$, $\Lambda$, and $\Xi$ as a function of $p_{T}/n_{q}$ in Au+Au collisions at $\sqrt{s_{NN}}$ = 200 GeV. Panel (b) : The $v_{2}$ of quarks as a function of  $KE_{T}^{q}$= $\sqrt{(p_{T}/n_{q})^{2} + m_{q}^{2}} -m_{q}$, where $p_{T}$ is the transverse momentum of hadron. Here we consider $m_{q}$ = 4 MeV/c$^{2}$ and 100 MeV/c$^{2}$ for light and strange quarks, respectively.  Panel (c) : Same as panel (b), but here $m_{q}$ = 140 MeV/c$^{2}$  for strange quarks. Blue dashed curves are the fit to light quarks $v_{2}$ using Eq.~\ref{eq5} and red dashed lines are the constant polynomial  fit to the ratios shown in the respective bottom panels.}
\label{fig2a}
\end{center}
\eef

\bef
\begin{center}
\includegraphics[scale=0.8]{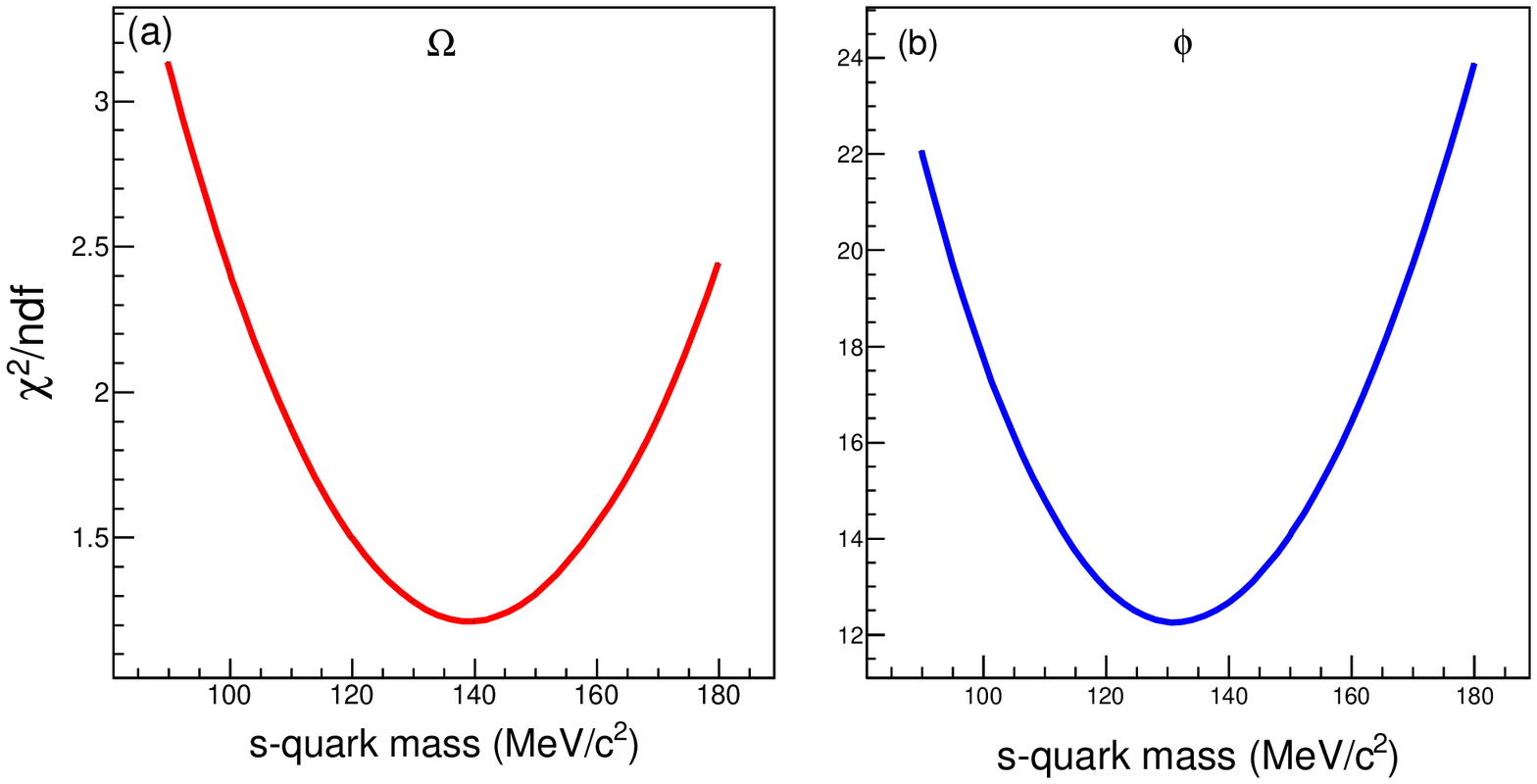}
\caption{(Color online)  $\chi^{2}/ndf$ for different s-quark mass for the best match between strange and light quarks $v_{2}$  as a function of transverse kinetic energy. This is for Au+Au collisions at $\sqrt{s_{NN}}$ =200 GeV. Panel (a): s-quarks $v_{2}$ derived from the $\Omega$ is used.  Panel (b): s-quarks $v_{2}$ derived from the $\phi$ is used. }
\label{fig2b}
\end{center}
\eef

The $v_{2}$ as a function of the transverse kinetic energy of particles produced in ultra-relativistic heavy-ion collisions is found to be a  more effective way of showing scaling among different hadrons, as shown by the PHENIX experiment~\cite{scaling_phenix}. It was shown that  $v_{2}/n_{q}$ as a function of $KE_{T}$/$n_{q}$ shows better scaling compared to $v_{2}/n_{q}$ as a function of $p_{T}/n_{q}$. 
Here $KE_{T}$ = $m_{T}$ -$m_{0}$ is the transverse kinetic energy of the hadrons. However, this scaling has no strong theory background. 

In this paper, we have shown an alternative empirical scaling relation between $v_{2}/n_{q}$ as a function of  $KE_{T}^{q}$ = $\sqrt{(p_{T}/n_{q})^{2} + m_{q}^{2}} -m_{q}$, where $p_{T}$ is the transverse momentum of hadron.
If we use  $m_{q}$ as the bare mass of a quark and $p_{T}/n_{q}$ as its transverse momentum, then the quantity $KE_{T}^{q}$ is equivalent to transverse kinetic energy of quark.  
 The values of bare or current masses of up ($u$), down ($d$) and strange ($s$) quarks are 1.8-2.8 MeV/$c^{2}$, 4.3-5.2 MeV/$c^{2}$ and 92-104 MeV/$c^{2}$, respectively~\cite{pdg}.  Since mass of $u$ and $d$ quarks are very close, in this study we took 4 MeV/$c^{2}$ as an average value of $u$ and $d$ quarks mass. Figure.~\ref{fig2a}  (middle panel), shows $v_{2}/n_{q}$ as a function of $KE_{T}^{q}$  using  mass of strange ($m_{s}$) quarks is equal to 100 MeV/$c^{2}$. We can see the $v_{2}/n_{q}$ as a function of $KE_{T}^{q}$  shown better scaling compared to $v_{2}/n_{q}$ as a function of   $p_{T}/n_{q}$ (left panel). 
However, we have seen the best scaling obtained when we use $m_{s}$ $\sim$ 130-140 MeV/$c^{2}$. For different $s$-quark mass, we have calculated $\chi^{2}/ndf$  to get the best match between strange and light quarks $v_{2}$ which is shown in Fig.~\ref{fig2b}. Since we have s-quark $v_{2}$ derived from both $\phi$ and $\Omega$, we have shown $\chi^{2}/ndf$ distribution for $\phi$ and $\Omega$ separately.  We can see from Fig.~\ref{fig2b}, that $s$-quark mass $m_{s}$ = 130-140 MeV/c$^{2}$ gives best $\chi^{2}/ndf$ while comparing with light quark $v_{2}$ as a function of transverse kinetic energy.
The value of  s-quark mass $m_{s}$ = 130-140 MeV/c$^{2}$ is purely empirical  and does not yet have a theoretical background. Hence, in this paper, we present the scaling relation using both $m_{s}$ = 100 MeV/c$^{2}$ (PDG value) and 140 MeV/c$^{2}$ (best $\chi^{2}/ndf$).

Further, to quantify the scaling, we have shown the ratio between light and strange quarks $v_{2}$ in the respective bottom panels. 
 To calculate the ratios, we have fitted the light quarks $v_{2}$ with the following function,
\begin{equation}
 v_{2}(x) = \frac{a n}{1+\exp(-(x n-b)/c)} - d n,
 \label{eq5}
 \end{equation}
where $a$, $b$, $c$,  $d$ and $n$ are the fit parameters.  The ratios are then fitted with a constant polynomial to get the average deviation ($p_{0}$) of the scaling.\\
In Fig.~\ref{fig2a}, the pion $v_{2}$ is not included since the $n_{q}$ scaled $v_{2}$ of pions is found to be higher than that of protons, which could be due to the effect of resonance decay in pion as reported in Ref.~\cite{coal_all_4}.

\subsection{The test of scaling at RHIC BES energies}
The validity of the observed scaling is examined at other BES energies.
Surprisingly, we find that such scaling works well even at different energies. Figure~\ref{fig3} shows $v_{2}/n_{q}$ as a function of $KE_{T}^{q}$ in 0-80\% minimum bias Au+Au collisions at $\sqrt{s_{NN}}$ = 39, 19.6, and 11.5 GeV. At lower beam energies, we have used anti-proton $v_{2}$ instead of proton to avoid the effect of transported quarks. We have seen the new proposed scaling  works equally good for both 39 and 19.6 GeV, for 11.5 GeV we need high statistics measurements to make any conclusion. The upcoming new data from RHIC Beam Energy Scan Phase-II (BES-II) will be useful to check this empirical scaling at lower energies, $\sqrt{s_{NN}}$ $<$19.6 GeV.

\bef
\begin{center}
\includegraphics[scale=0.7]{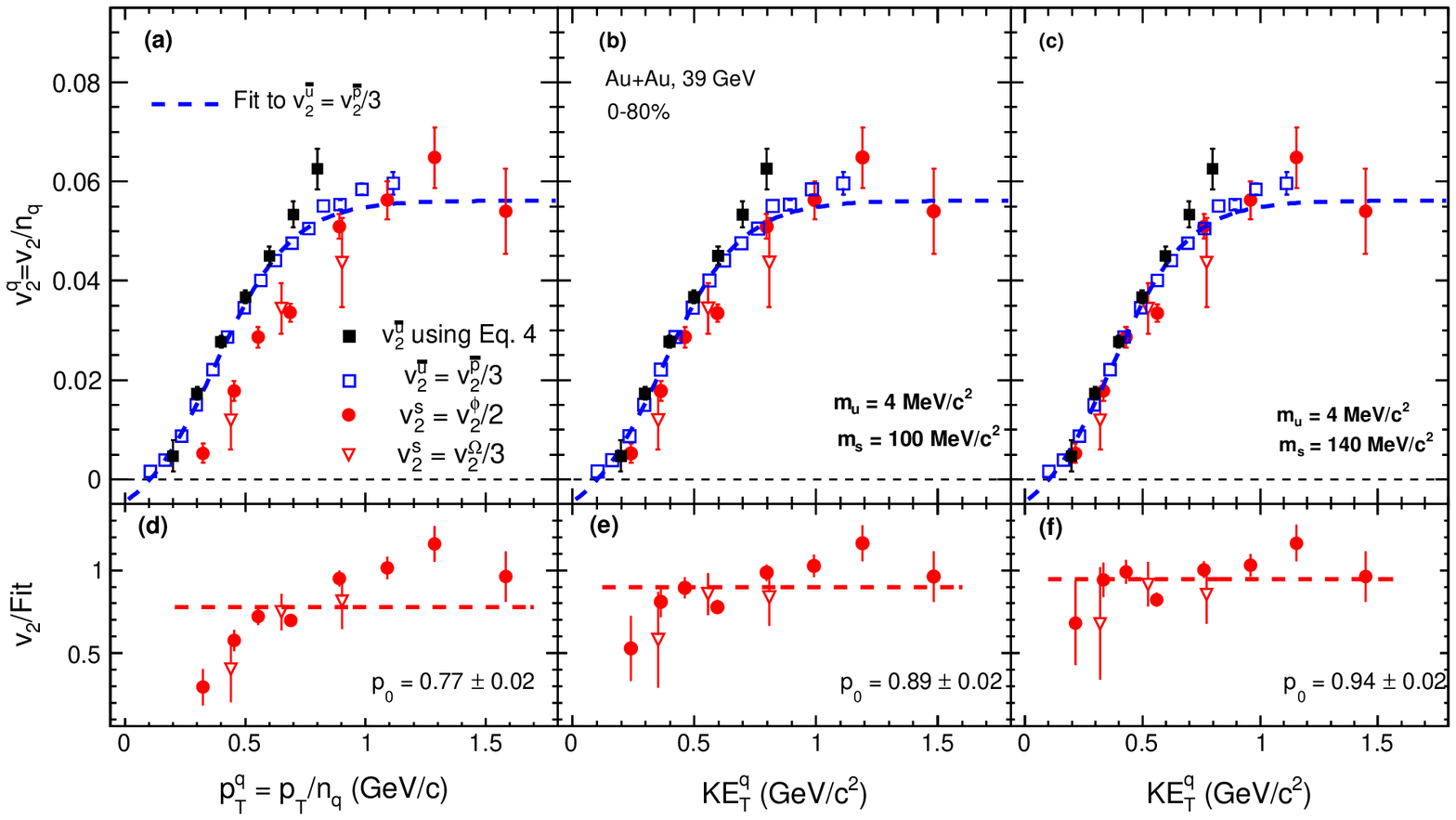}
\includegraphics[scale=0.7]{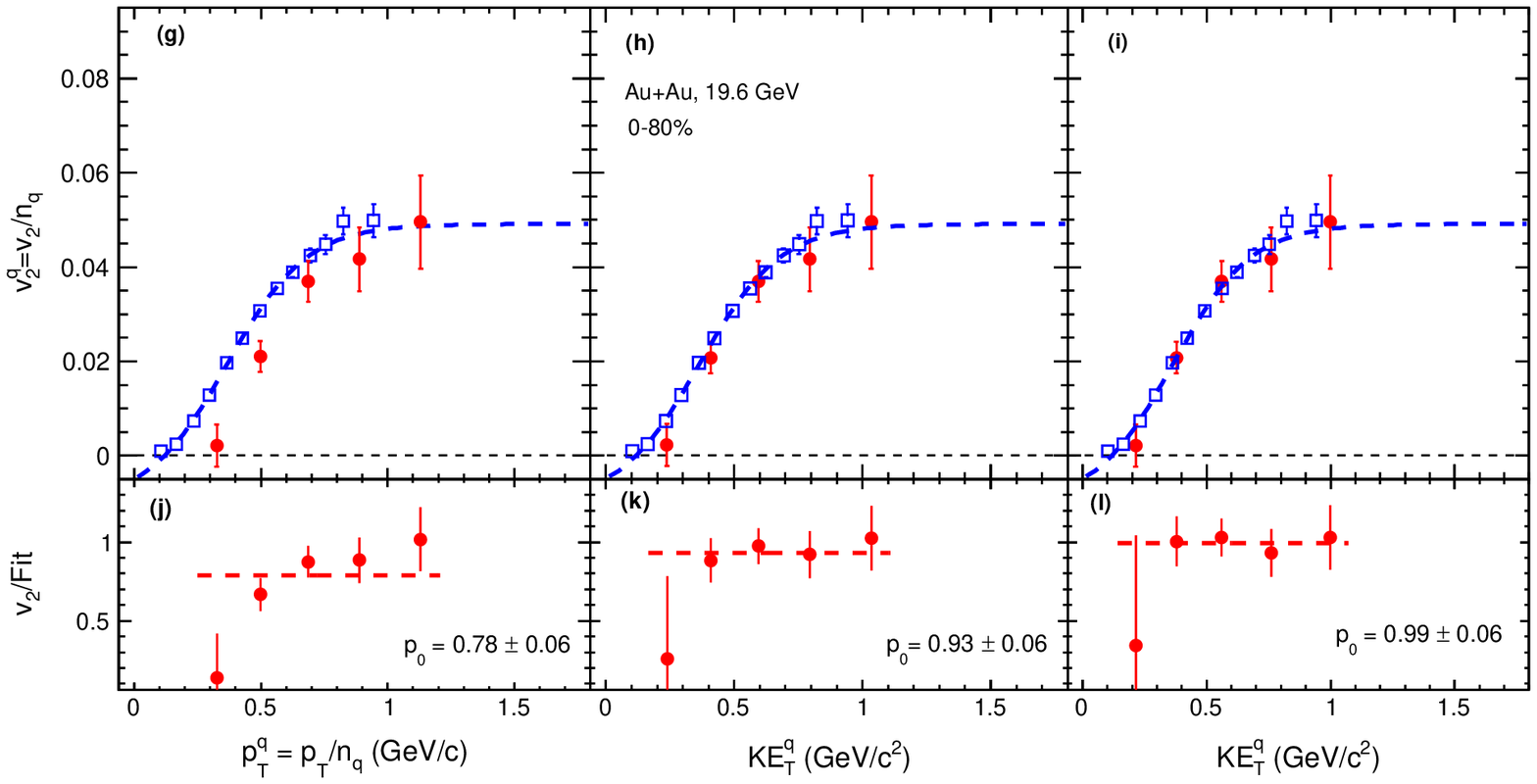}
\includegraphics[scale=0.7]{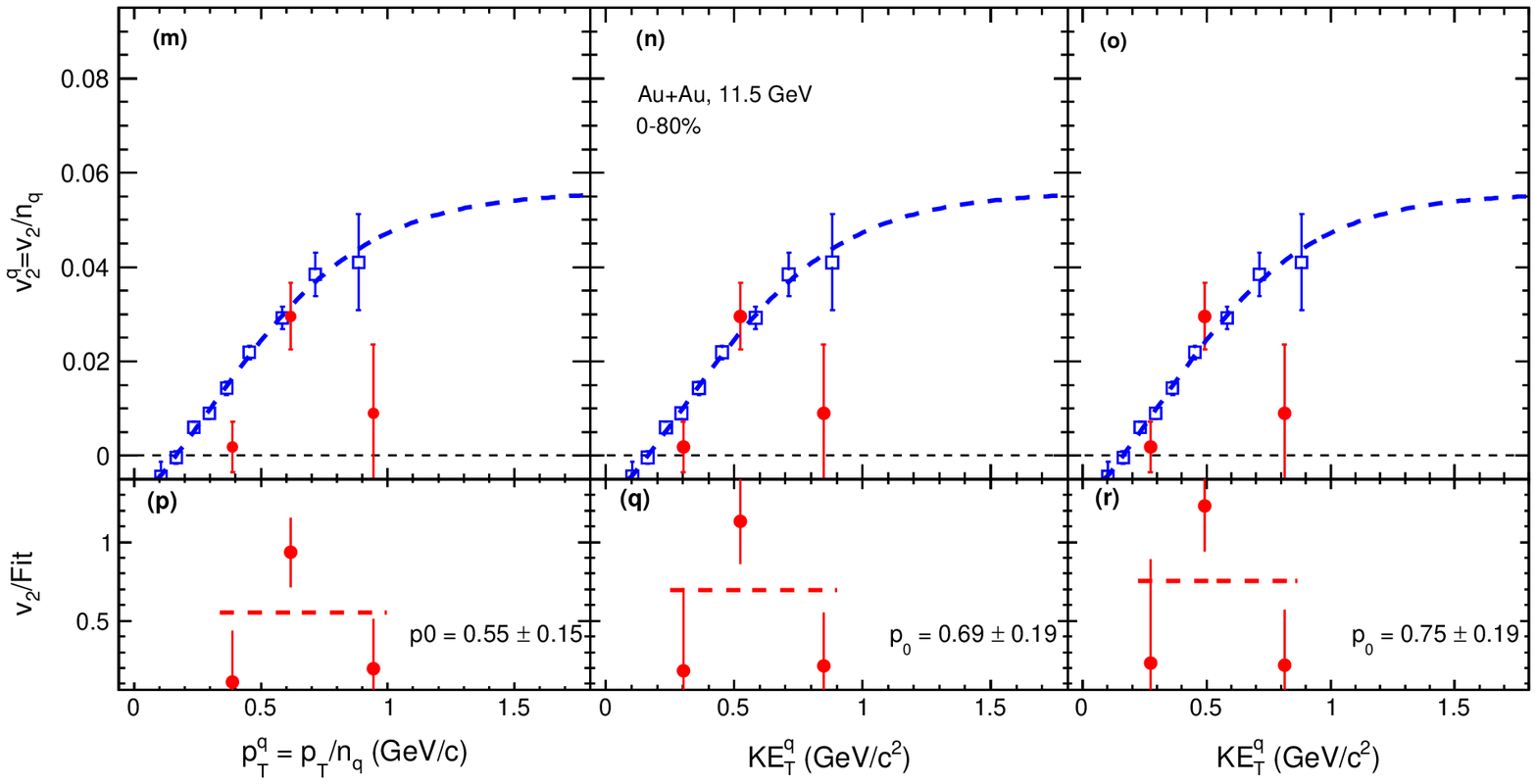}
\caption{(Color online) The left panel : The  $v_{2}$ of light and strange quarks obtained from $\bar{p}$ and $\phi$ (and $\Omega$, $\bar{\Lambda}$, and $\bar{\Xi}$ for 39 GeV only) as a function of $p_{T}/n_{q}$ in 0-80\% central Au+Au collisions at $\sqrt{s_{NN}} $ = 39, 19.6 and 11.5 GeV~\cite{pidv2_bes1,pidprl_bes1}. The middle panel: The $v_{2}$ of quarks as a function of  $KE_{T}^{q}$= $\sqrt{(p_{T}/n_{q})^{2} + m_{q}^{2}} -m_{q}$, where $p_{T}$ is the transverse momentum of hadron. Here we consider $m_{q}$ = 4 MeV/c$^{2}$ and 100 MeV/c$^{2}$ for light and strange quarks, respectively. The right panel: same as  the middle panel , but   $m_{q}$ = 140 MeV/c$^{2}$  for strange quarks.  Blue dashed curves are the fit to light quarks $v_{2}$ using Eq.~\ref{eq5} and red dashed lines are the constant polynomial  fit to the ratios shown in the respective bottom panels. }
\label{fig3}
\end{center}
\eef

\subsection{The test of scaling at LHC energy}
We have tested the scaling at LHC energy, where we have used the latest measurement of $\phi$ and proton $v_{2}$ at $\sqrt{s_{NN}} $ = 5.02 TeV by ALICE experiments~\cite{alice_502}. We do not have data for $\Omega$ and $\Xi$ at this energies.  Figure~\ref{fig4} shows $v_{2}/n_{q}$ as a function of $p_{T}/n_{q}$ and $KE_{T}^{q}$ in 20-30\% Pb+Pb collisions at $\sqrt{s_{NN}} $= 5.02 TeV.  We can see that $v_{2}/n_{q}$ for  $\phi$ meson is lower than that of protons at the low $p_{T}$ region when plotted as a function of  $p_{T}/n_{q}$. However,  the $v_{2}/n_{q}$ as a function of $KE_{T}^{q}$ shows similar values of $v_{2}$ for both proton and $\phi$.  This is true for other centrality collisions as well (not shown here).
 It is important to mention that the empirical scaling of $v_{2}/n_{q}$  as a function of $KE_{T}^{q}$ works well at both RHIC and LHC energies, where center-of-mass energies are very different, LHC is a factor 100 times higher than RHIC BES energies. This observation is interesting and needs further theoretical studies to understand the true mechanism behind the observed scaling in the measured hadrons $v_{2}$.

\subsection{The test of scaling in higher order flow harmonics}
In the end, we performed another systematic check using triangular flow harmonic ($v_{3}$). It was shown before that, in case of $v_{3}$, one needs to divide the $v_{3}$ by $n_{q}^{3/2}$ to see the scaling~\cite{XuSun}.
Figure~\ref{fig5} (a) shows  $v_{3}/n_{q}^{3/2}$ as a function of  $p_{T}/n_{q}$ for proton and $\phi$ measured by the STAR experiment in Au+Au collisions at $\sqrt{s_{NN}}$  = 200 GeV~\cite{v3_200}. Like $v_{2}$, we do see the scaled $v_{3}$ of $\phi$ is lower than scaled proton $v_{2}$ at low $p_{T}$. However, if we plot   $v_{3}/n_{q}^{3/2}$ as a function of $KE_{T}^{q}$, we do see good consistency as observed for $v_{2}$.

\bef
\begin{center}
\includegraphics[scale=0.8]{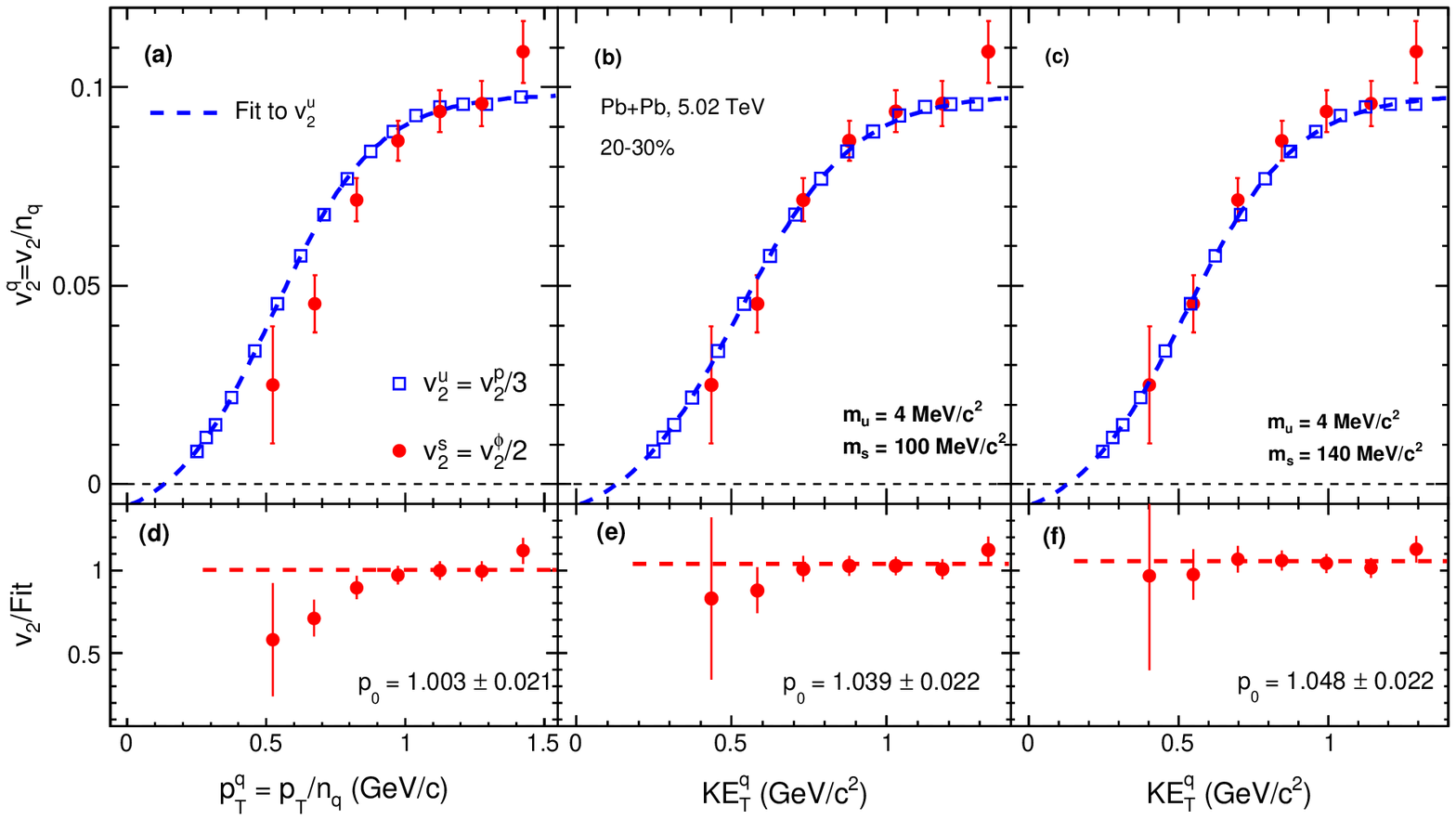}
\caption{(Color online) Panel (a) :  The  $v_{2}$ of light and strange quarks obtained from proton, and $\phi$ as a function of $p_{T}/n_{q}$ in  20-30\% central Pb+Pb collisions at $\sqrt{s_{NN}} $ = 5.02 TeV~\cite{alice_502}. Panel (b) :  The $v_{2}$ of quarks as a function of  $KE_{T}^{q}$= $\sqrt{(p_{T}/n_{q})^{2} + m_{q}^{2}} -m_{q}$, where $p_{T}$ is the transverse momentum of hadron. Here we consider $m_{q}$ = 4 MeV/c$^{2}$  and 100 MeV/c$^{2}$  for light and strange quarks, respectively. Panel (c) : Same as panel (b), but here  $m_{q}$ = 140 MeV/c$^{2}$   for strange quarks. Blue dashed curves are the fit to light quarks $v_{2}$ using Eq.~\ref{eq5} and red dashed lines are the constant polynomial  fit to the ratios shown in the respective bottom panels. }
\label{fig4}
\end{center}
\eef

\bef
\begin{center}
\includegraphics[scale=0.8]{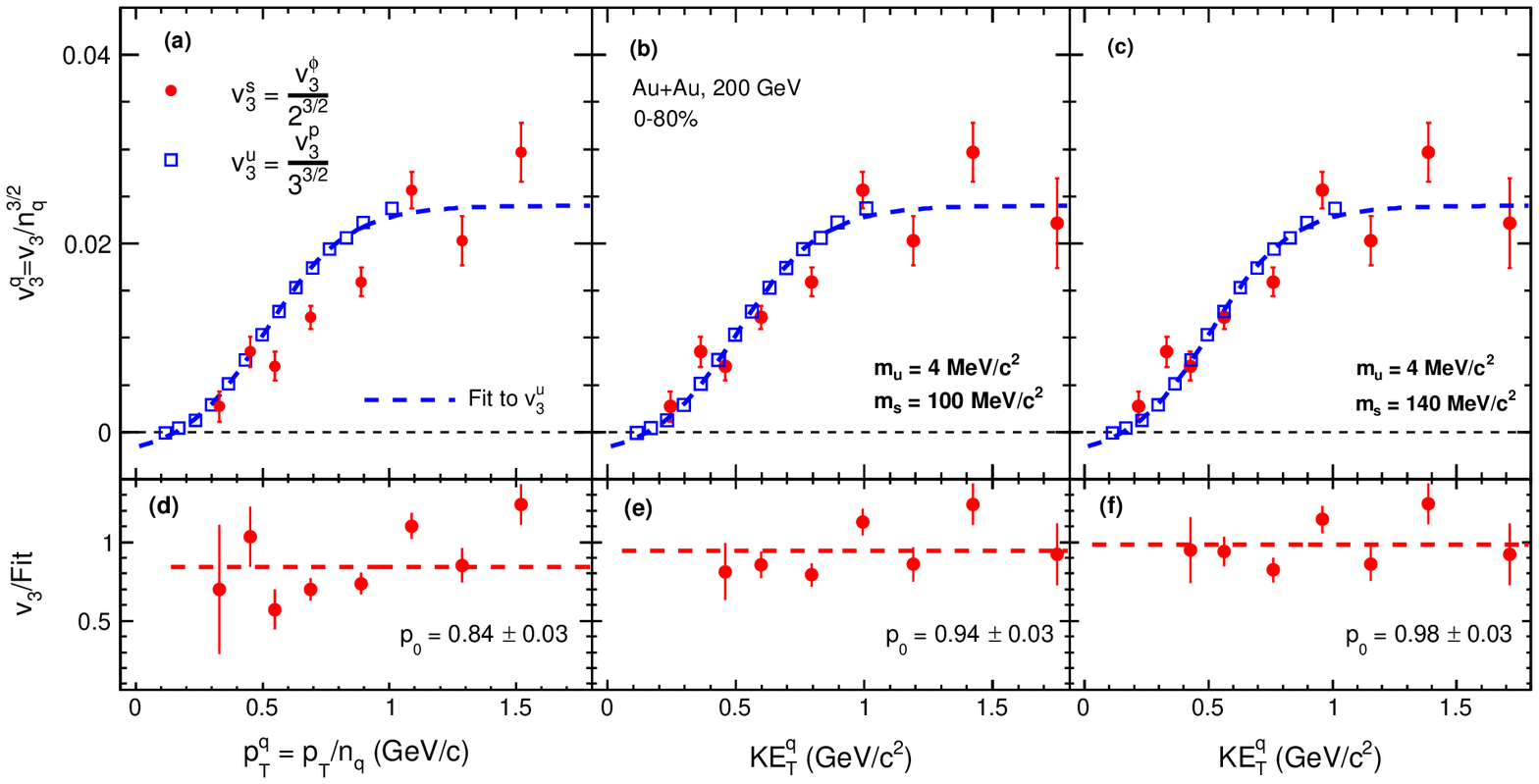}
\caption{(Color online) Panel (a): The  $v_{3}/n_{q}^{3/2}$ of proton and $\phi$ as  a function of $p_{T}/n_{q}$ in Au+Au collisions at  $\sqrt{s_{NN}} $ = 200 GeV~\cite{v3_200}. Panel (b):  $v_{3}/n_{q}^{3/2}$ of proton and $\phi$  a function of  $KE_{T}^{q}$= $\sqrt{(p_{T}/n_{q})^{2} + m_{q}^{2}} -m_{q}$, where $p_{T}$ is the transverse momentum of hadron. Here we consider $m_{q}$ = 4 MeV/c$^{2}$ and 100 MeV/c$^{2}$ for light and strange quarks, respectively.  Panel (c) : Same as panel (b), but  here  $m_{q}$ = 140 MeV/c$^{2}$  for strange quarks. Blue dashed curves are the fit to light quarks $v_{3}$ using Eq.~\ref{eq5} and red dashed lines are the constant polynomial  fit to the ratios shown in the respective bottom panels.}
\label{fig5}
\end{center}
\eef

\section{Summary}
In summary, a compilation of the available data at RHIC and LHC energies for the elliptic flow and triangular flow of identified hadrons is presented.
Within the framework of the quark recombination model, $v_{2}$ of light and strange quarks are obtained from the measured $v_{2}$ of identified hadrons. 
We find that the $v_{2}$ of strange quarks obtained from $\phi$-meson $v_{2}$ is found to be similar to that of from $\Omega$ $v_{2}$ in Au+Au collisions at $\sqrt{s_{NN}}$ = 200 GeV, indicating that both $\phi$ and $\Omega$ are produced through quark recombination at top RHIC energy. However, we find the $v_{2}$ of strange quarks is consistently smaller than that of light quarks at the low transverse momentum. 
We have shown an empirical relation between $v_{2}$ of light and strange quarks, where   $v_{2}$ of quarks scales as a function of  the transverse kinetic energy of quarks. The transverse kinetic energy of quarks has been calculated using the bare mass of quarks.  Such empirical relation holds for both RHIC and LHC energies. It is shown that $v_{3}$ of the measured hadrons also follows the same scaling.  The observation of such new scaling in flow measurement  is purely empirical and needs further theoretical investigation.



\end{document}